\documentclass[aps,prd,preprint,superscriptaddress,tightenlines,nofootinbib]{revtex4}

\usepackage{graphicx}
\usepackage{dcolumn}
\usepackage{bm}

\newcommand{\gev}{\,\mbox{GeV}}
\newcommand{\mev}{\,\mbox{MeV}}
\newcommand{\invpb}{\,\mbox{pb}^{-1}}
\newcommand{\diel}{e^+e^-}
\newcommand{\dimu}{\mu^+\mu^-}
\newcommand{\dipion}{\pi^+\pi^-}
\newcommand{\dilep}{\ell^+\ell^-}
\newcommand{\jpsi}{J/\psi}
\newcommand{\bll}{{\cal B}(\jpsi \to \dilep)}
\newcommand{\jpsitoany}{J/\psi \to X }
\newcommand{\pp}{\psi(2S)}

\begin{document}

\preprint{CLNS 05/1910}       
\preprint{CLEO 05-5}         

\title{Measurement of the Branching Fractions for $\jpsi\to\ell^+\ell^-$} 

\author{Z.~Li}
\author{A.~Lopez}
\author{H.~Mendez}
\author{J.~Ramirez}
\affiliation{University of Puerto Rico, Mayaguez, Puerto Rico 00681}
\author{G.~S.~Huang}
\author{D.~H.~Miller}
\author{V.~Pavlunin}
\author{B.~Sanghi}
\author{E.~I.~Shibata}
\author{I.~P.~J.~Shipsey}
\affiliation{Purdue University, West Lafayette, Indiana 47907}
% ...
\author{(CLEO Collaboration)} 
\noaffiliation

\date{March 10, 2005}

\begin{abstract} 
We present measurements of the branching fractions for 
$\jpsi \to \diel$ and $\dimu$ using 3M~$\pp$~decays 
collected with the CLEO detector operating
at the CESR $\diel$ collider. We obtain 
${\cal B} (\jpsi \to \diel) = ( 5.945 \pm 0.067 \pm 0.042 )\%$ and 
${\cal B} (\jpsi \to \dimu) = ( 5.960 \pm 0.065 \pm 0.050 )\%$, leading
to an average of 
${\cal B} (\jpsi \to \dilep) = ( 5.953 \pm 0.056 \pm 0.042 )\%$ and 
a ratio of
${\cal B} (\jpsi \to \diel) / {\cal B} (\jpsi \to \dimu) = 
( 99.7 \pm 1.2 \pm 0.6 )\%$, 
all consistent with, but more precise than, previous measurements.
\end{abstract}

\maketitle

The $\jpsi$~meson is often experimentally identified through its two
largest and cleanest exclusive decay modes, $\jpsi \to \diel$ or $\dimu$, 
and hence the corresponding branching fractions
are of general interest.
The process is thought to occur through
annihilation of the $c \bar c$~pair into a virtual photon
which then materializes as a lepton pair, thereby
relating to the $c \bar c$~wave function overlap at the origin
and playing a direct role in potential models~\cite{potentials}.
The dilepton branching fraction serves as 
an ingredient in the measurement~\cite{babar} of the $J/\psi$ dileptonic
and total widths ($\Gamma_{ee}$ and $\Gamma_{\rm tot}$). It also acts as
a normalization in the
comparison of $\psi(2S)$ and $\jpsi$ exclusive final
state production of light hadrons;
the assumption is that the underlying hard reaction, namely
annihilation of the heavy quark pair, is the same in both cases.

The current experimental status is that both lepton pair species
have been measured to be equal in production rate,
as expected from lepton universality (in combination with
a negligible correction for phase space), 
at branching fractions of $\sim 5.9\%$.
A relative precision of $1.7\%$ on each of ${\cal B} (\jpsi \to \diel)$
and ${\cal B} (\jpsi \to \dimu)$ has been achieved
through an average~\cite{pdg2004} over measurements,
which is dominated by a result from BES~\cite{bes_jpsidilep}:
${\cal B} (\jpsi \to \diel)  = ( 5.90 \pm 0.05 \pm 0.10 ) \%$,
${\cal B} (\jpsi \to \dimu)  = ( 5.84 \pm 0.06 \pm 0.10 ) \%$,
and
${\cal B} (\jpsi \to \dilep) = ( 5.87 \pm 0.04 \pm 0.09 ) \%$.
The current precision of $\bll$ is a significant
contributor to the uncertainties on $\Gamma_{ee}$ and 
$\Gamma_{\rm tot}$ obtained from measurement
of radiative return ($e^+e^-\to \gamma J/\psi$, $J/\psi\to\mu^+\mu^-$)
cross sections~\cite{babar}.

In this article we describe measurements of ${\cal B}(\jpsi\to\diel)$
and ${\cal B}(\jpsi\to\dimu)$ using 
the decay $\psi(2S) \to \pi^+\pi^-\jpsi$. 
The experimental procedure is straightforward and
consists of determining the ratios of the 
numbers of exclusive $\jpsi \to \dilep$
decays for $\ell = e$ and $\mu$, $N_{\diel}$ and $N_{\dimu}$, to 
the number of inclusive $\jpsitoany$ decays, $N_{X}$, where
$X$ means all final states.
The branching fractions 
will be calculated as
$
{\cal B}(\jpsi \to \dilep) = 
   {(N_{\dilep}/\epsilon_{\dilep})}/{(N_X/\epsilon_X)},
$
where $\epsilon_{\dilep}$ and $\epsilon_X$ represent the
detection probabilities for the exclusive and inclusive events,
respectively.

We use $e^+e^-$~collision data at and below the $\pp$~resonance,
$\sqrt s = 3.686\gev$ (${\cal L} = 5.86\invpb$) and
$\sqrt s = 3.670\gev$ (${\cal L} = 20.46\invpb$),
collected with the CLEO detector~\cite{CLEOdetector}
operating at the Cornell Electron Storage Ring (CESR)~\cite{cesr}. 
The CLEO detector features a solid angle coverage of $93\%$ for
charged and neutral particles. 
The charged particle system operates in a 1.0~T~magnetic field
along the beam axis and achieves a momentum resolution of
$\sim 0.6\%$ at momenta of $1\gev/c$. 

Identification of $\dipion\jpsi$ candidates is performed
by tagging the dipion pair:
two tracks of opposite charge with $m(\dipion) = 400-600\mev$, 
$| \cos \theta | < 0.83$ (where $\theta$ is the polar angle
of each track
with respect to the $e^+$~direction), and, to avoid
tracks which bend back into the tracking
detectors before they can enter the calorimeter (`curlers'),
a momentum component transverse to the beam axis exceeding~$150\mev/c$. 
The number of $\jpsitoany$ events is determined
from a fit to the distribution of the invariant mass
recoiling against the dipion pair, $m$($\dipion$-recoil),
in the $\pi^+\pi^-X$~sample after applying this preselection.

To select event samples of $\jpsi\to\dilep$, we demand that 
candidate events fulfil the following requirements:  
The lepton pair, consisting of the two highest-momentum
tracks in the event, must satisfy the very loose identification
criteria of $E/p > 0.85$ for one electron
and $E/p>0.5$ for the other, or $E/p < 0.25$ and $E/p<0.5$
in case of muons, where $E$ is the measured calorimetric energy
deposition of each track and $p$ is its measured momentum. 
The invariant mass of the track pair must be consistent
with that of a $J/\psi$, with $m(\dilep) = 3.02 - 3.22\gev$. 
In order to salvage lepton pairs that
have radiated photons and would hence fail
the $\jpsi$~mass cut, we add bremsstrahlung photon candidates
found within a cone of $100\,\mbox{mrad}$ to the track three-vector at
the $\diel$~interaction point. We impose loose restrictions on 
the absolute momentum and energy of the event:
$( E_{\jpsi} + E_{\dipion} ) / \sqrt s = 0.95 - 1.05$, 
$ | |p_{\jpsi}| - | p_{\dipion} | | / \sqrt s < 0.07$.
We search for the same signature in data 
taken $15\mev$ below the $\psi(2S)$ resonance and find a 
level of population consistent with 
the Breit-Wigner tail of the $\psi(2S)$. 
Backgrounds from other $\psi(2S)$ decays mimicking 
the desired signature are subtracted, which is a relative 
reduction of 0.07\% for dielectrons (mostly $\pp \to 
\eta\jpsi,\ \jpsi\to\diel$) and 0.2\% for dimuons (consisting
of $\pp \to \eta\jpsi,\ \jpsi\to\dimu$ at a similar level 
as electrons, as well as $\pp \to \dipion$ and to a lesser 
extent $\pp \to \rho\pi$). The resulting event yield is~$N_{\dilep}$.    

The detection probabilities are determined from MC~simulation 
using the {\tt EvtGen} generator~\cite{evtgen} and a 
GEANT-based~\cite{geant} detector simulation.
The dipion invariant mass distribution 
as produced by {\tt EvtGen} is slightly suppressed at high 
and low $m(\pi^+\pi^-)$ to better match the data, 
altering the efficiencies by $<0.5\%$.

\begin{figure}
\includegraphics*[width=6.5in]{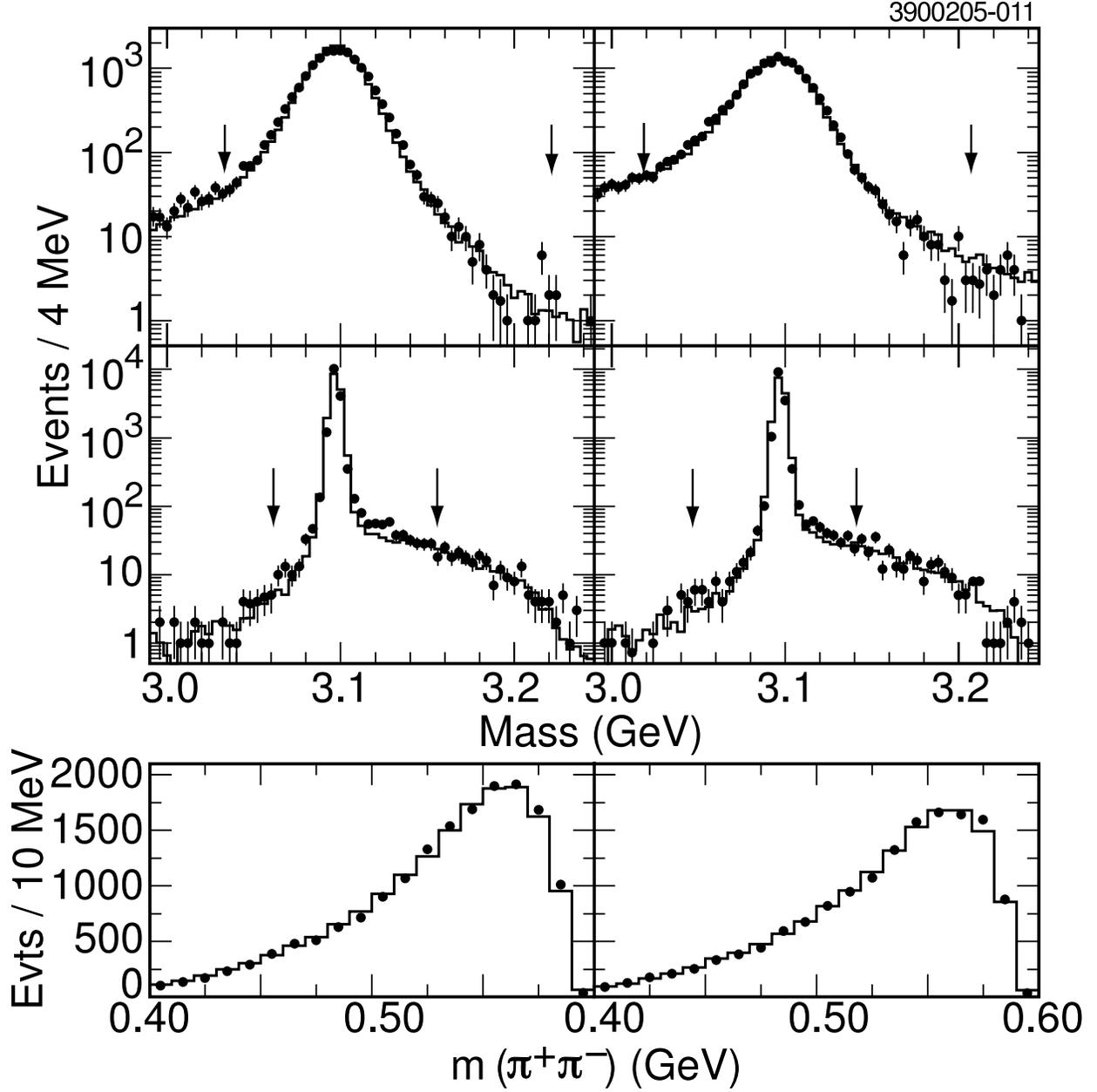}
\caption{
For $\psi(2S)\to\pi^+\pi^-\ell^+\ell^-$
dimuon (left) and dielectron (right)
candidate events in the $\psi(2S)$
data (solid circles), MC simulation of signal (solid histogram),
distributions of the dilepton mass (top),
the mass recoiling against the $\pi^+\pi^-$ pair (middle),
and the dipion invariant mass.
The arrows
shown in each plot indicate nominal cut values, 
which are applied for the other plots in the figure. 
}
\label{fig:pi+pi-_minv}
\end{figure}

\begin{figure}
\includegraphics*[width=6.5in]{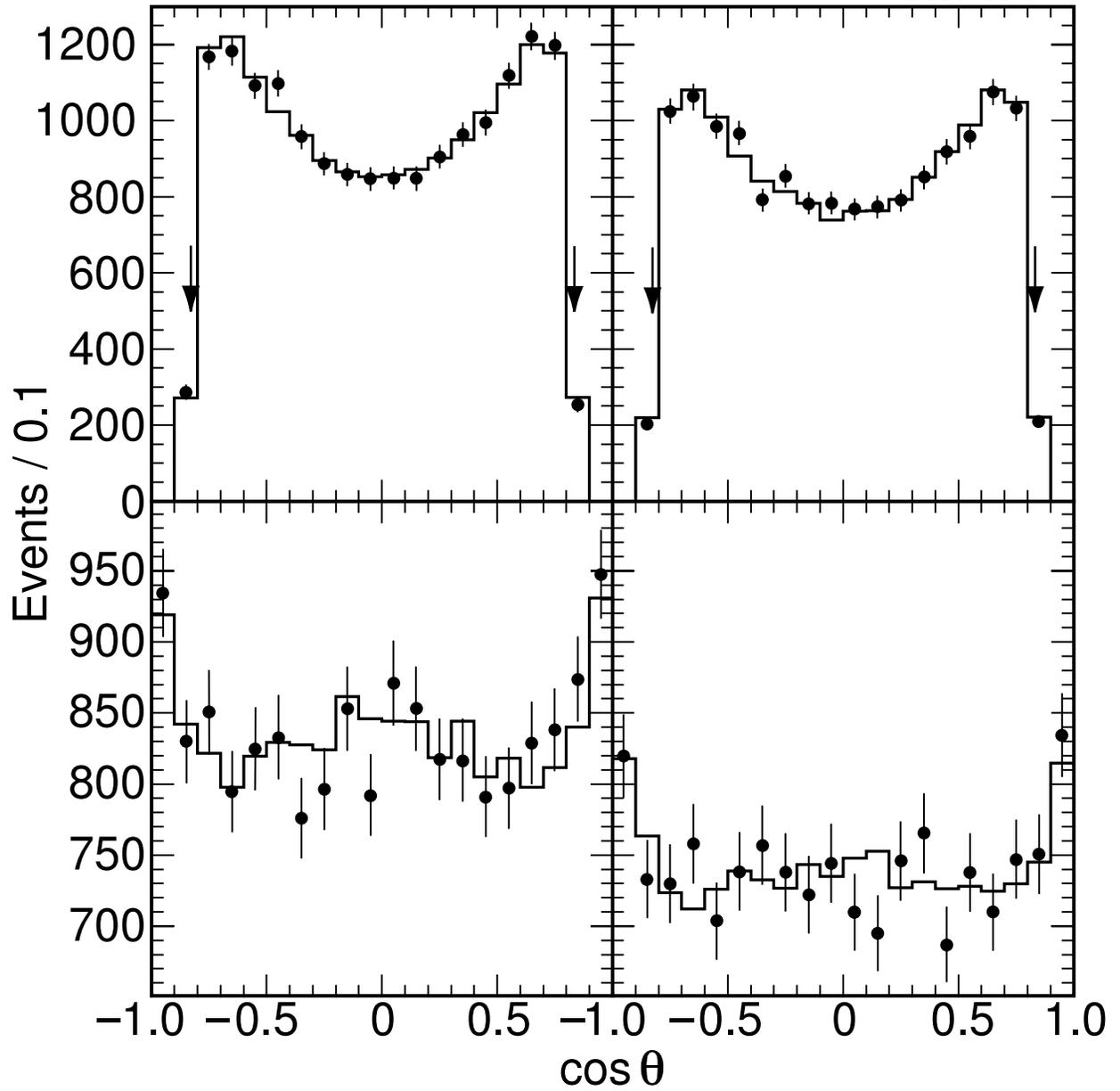}
\caption{
For $\psi(2S)\to\pi^+\pi^-\ell^+\ell^-$
dimuon (left) and dielectron (right)
candidate events in the $\psi(2S)$
data (solid circles) and MC simulation of signal 
(solid histogram), 
the polar angles of the positively charged lepton~(top)
and of the $\jpsi$~(bottom).       
}
\label{fig:pi+pi-_ang}
\end{figure}

We demonstrate the statistical power of the data sample,
its cleanliness, and the excellent agreement observed
between data and Monte Carlo (MC) in Figures~\ref{fig:pi+pi-_minv} 
and~\ref{fig:pi+pi-_ang}, where we show the invariant masses of
the dipion pair (direct and recoil) and the dilepton pairs,
and also the lepton and $\jpsi$ polar angle distributions. 
Further evidence for the high degree of quantitative understanding
of $\psi(2S)$ decays to final states with a $J/\psi$ in CLEO can be found
in Ref.~\cite{xjpsi}.

\begin{figure}[thp]
\caption{Fit result (solid histogram) 
of the $\pi^+\pi^-$ recoil mass spectrum
in data (solid circles)
as described in the text.
The dashed curve represents the background shape.
\label{fig:jpsibr_fit} }
\includegraphics*[width=6.5in]{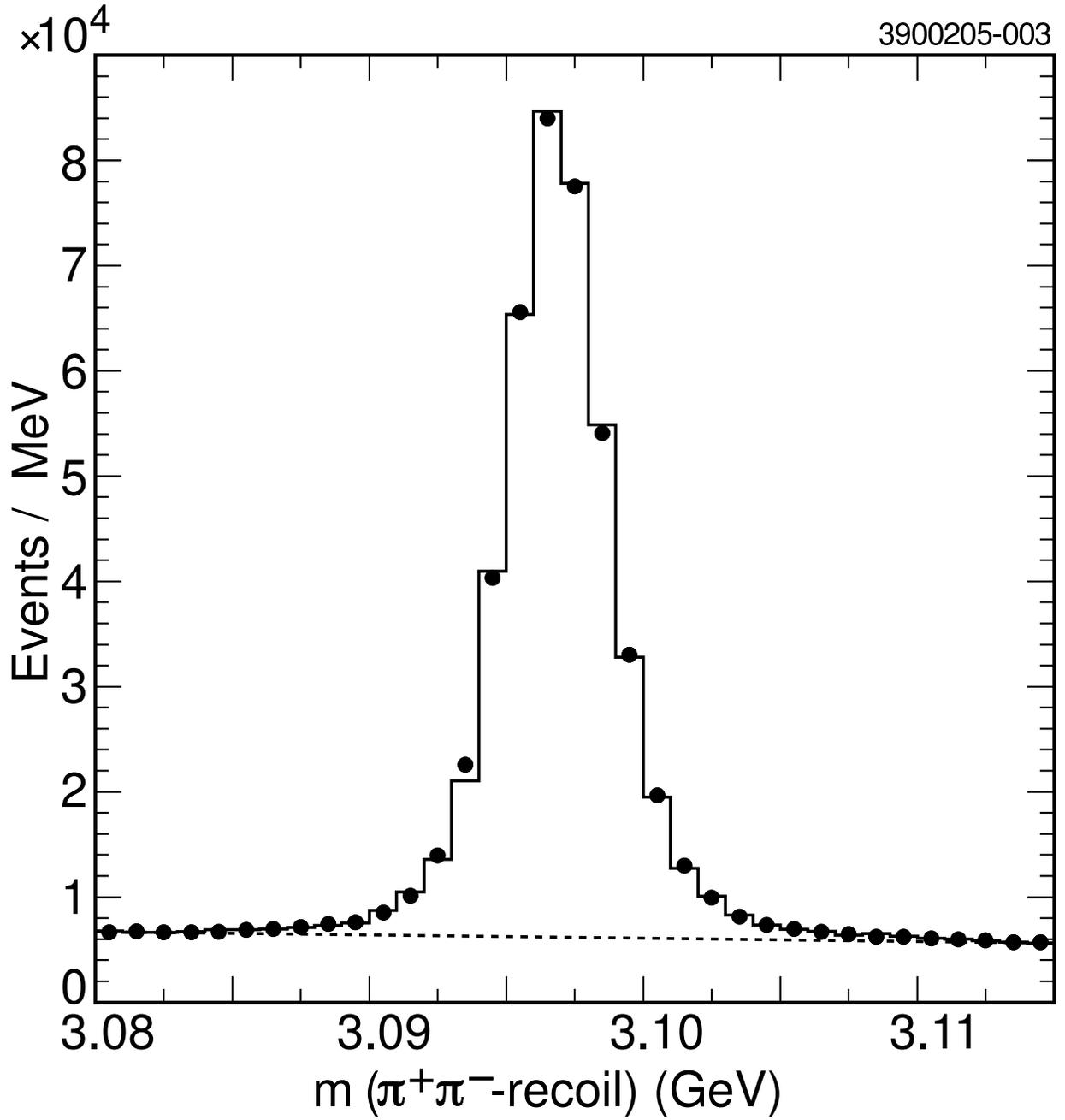}
\end{figure}

We follow a procedure similar to that employed 
by BES~\cite{bes_jpsidilep}  
and Mark~III~\cite{MK3_jpsidilep} to obtain the
raw number of $\jpsitoany$ decays, in which the 
$m$($\pi^+\pi^-$-recoil) 
spectrum is fit to obtain the number of $\jpsi$
candidates. 
Since the dipion emission occurs 
independently of the subsequent $\jpsi$~decay, the dipion
recoil mass shape can be taken from any cleanly determined
$\jpsi$~decay. This grants us considerable
freedom from the accuracy of MC in modeling
the momentum resolution.  We use the sum of
$\jpsi\to\diel$ and $\jpsi \to \dimu$, which is almost
background-free, for the signal shape of the dipion
recoil mass distribution.
As our MC does not perfectly describe the shape of data,
the $\pi^+\pi^-X$~data recoil mass distribution is fit with the
$\pi^+\pi^-\ell^+\ell^-$~signal shape from data. 
The fit, shown in Figure~\ref{fig:jpsibr_fit}, uses a  
second-order polynomial background shape.
The confidence level of the fit is 23\%.
Increasing the order of the polynomial describing
the background does not alter the fitted 
signal normalization nor substantially improve the fit.

Since the sum of known exclusive $\jpsi$
partial widths is small compared to the total
width, a MC sample must be chosen somewhat
arbitrarily to represent all $\jpsi$ decays and from which 
to obtain $\epsilon_X$.
We calculate the inclusive $\dipion\jpsi$, $\jpsi \to X$ 
counting efficiency for a selection of modes with
different charged and neutral multiplicities.
The efficiencies thus determined
are $\sim$40\% and vary by only $\sim$2\% (relative)
from low to high multiplicities, nearly 
an order of magnitude smaller variation than that reported 
in~\cite{bes_jpsidilep}. We attribute this effect to the finer 
segmentation in the CLEO tracking system~\cite{CLEOdetector}
relative to that of BES~\cite{BESIdetector}
and the consequent robustness of track-finding
in the presence of many charged particles.
The efficiency does exhibit a small dependence not 
only on the charged multiplicity, but also on the
neutral multiplicity. The addition of neutral particles
in the $\jpsi$ decay softens the momentum spectrum
of the charged tracks, causing some to be lost at low
momentum or small polar angles, and 
also adds to the track multiplicity through photon
conversions in the material of the inner detectors.
Curlers are also produced more often, which can
make pattern recognition more difficult. Even so,
such deleterious effects are very small.

\begin{figure}
\caption{Track multiplicity distribution
for $J/\psi$ decays produced in $\pp \to \pi^+\pi^-\jpsi$. 
Left: Signal MC for nine decay modes, right:
data distribution obtained from the sideband-subtracted
inclusive $\pi^+\pi^-$ samples (solid circles) 
and the fit to MC samples of different multiplicities (solid line). 
\label{fig:mulFit} }
\includegraphics*[width=6.5in]{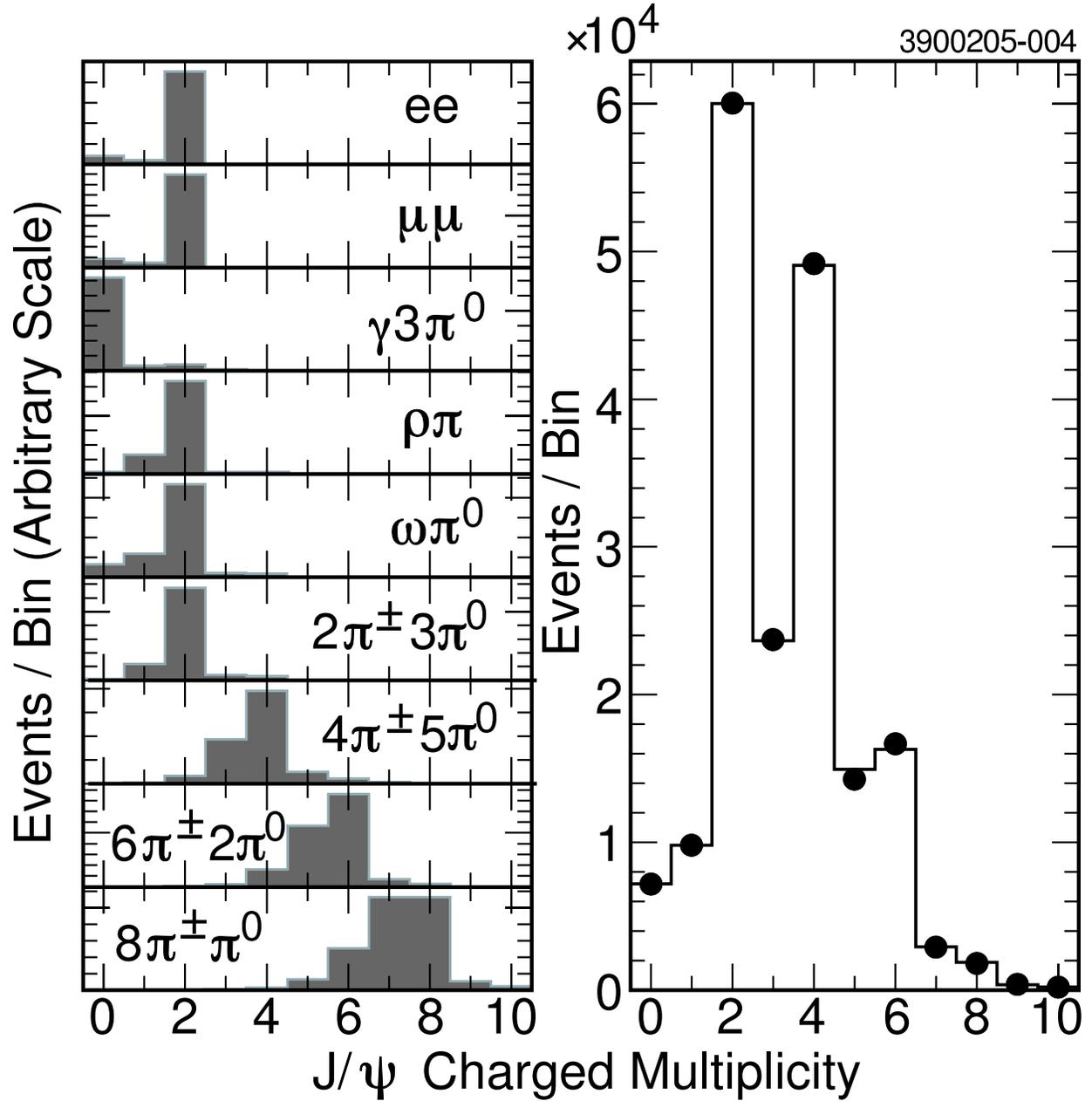}
\end{figure}

Given that the detection efficiency for $\jpsi$~decay
products in $\pi^+\pi^-\jpsi$~events
depends on the track multiplicity, we let
this quantity guide us in the choice of an appropriate
mixture of $\jpsi$~decay modes in MC (a basis set).
The measured
charged multiplicity, shown in Fig.~\ref{fig:mulFit}, is obtained
from events with $m$($\dipion$-recoil)=$3.090-3.104\gev$ 
after subtracting sideband contributions with 
$m$($\dipion$-recoil)=$[3.078-3.085]$, $[3.110-3.117]\gev$.
We investigate a variety of alternatives for the basis set.
The best fit to the $\jpsi$ charged multiplicity distribution
is obtained by incorporating 
three well-measured contributions~$i$,
$i=\diel$, $\dimu$, $\rho\pi$,
with their branching fractions~\cite{pdg2004} as fixed 
weights~$w_i$, and additional contributions 
with floating normalizations:
$\omega\pi^0\pi^0\to\gamma 3\pi^0$, 
$\omega\pi^0$, $2\pi^\pm 3\pi^0$, $4\pi^\pm 5\pi^0$,
$6\pi^\pm 2\pi^0$, $8\pi^\pm 1\pi^0$.
The result of this fit is an
effective branching fraction for each of the modes with floating
normalization. The only
purpose of this basis set is to reproduce the charged multiplicity
of the data, thereby permitting an accurate determination of~$\epsilon_X$;
the basis set is not intended to characterize exclusive
$\jpsi$ decays. Substituting for the basis set members
of multiplicities~2, 4, or~6 a similar mode with even one more or fewer~$\pi^0$
results in poorer representations
of the data. 

The inclusive efficiency $\epsilon_X$ 
is determined by fitting the $m$($\dipion$-recoil)
distribution obtained from mixing the MC events
in the proportions given by the weights in Table~\ref{tab:mulFit},
using a signal shape from $\dipion\dilep$~MC events.
The same value is obtained if, instead of
a fit to the weighted MC components, we calculate
a weighted average of the individual mode efficiencies.

The resulting raw and efficiency corrected yields are listed
in Table~\ref{tab:Sum}.

The fit result, $N_X$,  has a relative statistical
uncertainty of 0.65\%, dominated by the statistical 
uncertainty on the number of events determining the 
signal input shape. 
We assign the following additional systematic uncertainties,
with the same values for the dimuon and dielectron samples: 
for dipion charged track multiplicity weighting and choice 
of basis set, 0.3\%, and for yield fit-window variation, 0.5\%.
The former is set by the variation induced by using
other combinations of exclusive
$J/\psi$ decays (basis sets) that still closely match the multiplicity distribution
of the data; the latter by variations of the window limits as low
as 3.04~GeV and as high as 3.15~GeV.

It is to be noted that systematic effects related to soft
pion tracking cancel in the ratios.
Systematic studies for detection of the lepton pair 
track candidates follow.

We select $\pi^+\pi^-\ell^\pm(\ell^\mp)$ events by 
requiring $m$($\dipion$-recoil)=3.05-3.15$\gev$
and only {\sl one} lepton satisfying
$|\cos\theta_\ell|<0.83$ and $p=1.35 - 1.85$\gev/$c$,
but more strongly identified 
($E/p > 0.85$ for electrons, 
and for muons $E/p<0.25$ and a penetration of more than three
absorption lengths in the CLEO muon detector).
Each event must have a missing momentum 
direction of $|\cos\theta_{\rm miss}| < 0.75$ and a 
missing-mass-squared of less than $0.2 \gev^2$. 
In these events, we proceed to search for  a second
lepton of opposite charge, $E/p > 0.5$ ($e$) or $E/p<0.5$ ($\mu$)
and momentum $p>0.8\gev$, that produces a dilepton mass of
$m(\dilep) = 3.02 - 3.22\gev$. The fraction
of events in which we fail to identify the second
lepton is compared between data
and MC. The relative failure rate discrepancy between data and MC is 
$(0.85\pm0.13)$\% for dimuon events and $(0.01\pm0.20)$\%
(statistical errors only) for dielectron events.
These measured data$-$MC differences are used to establish 
yield correction factors
($0.995$ for muons, 1.000 for electrons) applied to $\epsilon_{\dilep}$  
and systematic 
uncertainties of $0.5\%$ ($0.2\%$) per $\dimu$ ($\diel$). 
Effects that can produce such a mismatch between data
and MC for either lepton species
include track reconstruction systematics and, more 
importantly, mismodeling of decay radiation, which leads to a loss
of events by causing a failure of the invariant mass requirement. 
The quoted uncertainty includes both.

The remaining source of systematic uncertainty
not addressed by the above lepton pair efficiency
study is modeling of the $E/p$ requirements,
which distinguish muons and electrons from each other
and from hadrons. This uncertainty is
determined by varying the value of the $E/p$ cuts
around the nominal values, and is found to be 
$0.1\%$ for both muon and electron pairs.

After including additional relative uncertainties from MC statistics (0.2\%)
and hadronic event trigger efficiency (0.2\%) in quadrature with the above,
the total relative
systematic error is $0.7\%$ for ${\cal B}(J/\psi\to e^+e^-)$
and $0.8\%$ for ${\cal B}(J/\psi\to\mu^+\mu^-)$.

Final results are 
${\cal B}(J/\psi\to e^+e^-)     = (5.945\pm0.067\pm0.042)\%$,
$ {\cal B}(J/\psi\to\mu^+\mu^-) = (5.960\pm0.065\pm0.050)\%$, and
${\cal B} (\jpsi \to \diel) / {\cal B} (\jpsi \to \dimu) 
                                = ( 99.7 \pm 1.2 \pm 0.6 )\%$.
Assuming lepton universality, the average is
${\cal B}(J/\psi\to \dilep)     = (5.953\pm 0.056 \pm 0.042)\%$, 
in which we have accounted for correlations among the errors.
These results are consistent with previous measurements, but 
improve considerably upon precision, constituting the 
most precise measurements to date. The 1.18\% relative (0.070\% absolute)
uncertainty on ${\cal B}(J/\psi\to \dilep)$ 
is significantly smaller than the previous smallest 
uncertainty~\cite{bes_jpsidilep}, allows
improvement in the precision of present and future
measurements~\cite{babar} of $\Gamma_{ee}$ and $\Gamma_{\rm tot}$,
and provides an important benchmark and calibration point
for potential models~\cite{potentials}.

We gratefully acknowledge the effort of the CESR staff 
in providing us with excellent luminosity and running conditions.
This work was supported by the National Science Foundation
and the U.S. Department of Energy.

\begin{table}[thp]
\setlength{\tabcolsep}{0.4pc}
\catcode`?=\active \def?{\kern\digitwidth}
\caption{
For different MC $\pi^+\pi^-J/\psi$, $J/\psi\to X$
decays (left column): 
The weights $w_i$ obtained in the fit to
the multiplicity, which are used in the fit 
extracting $\epsilon_X$, and the relative
difference between the efficiencies $\epsilon_X$,
obtained using the default $w_i$, and $\epsilon_i$,
employing the dipion recoil mass distribution from 
each channel alone. 
}
\label{tab:mulFit}
\begin{center}
\begin{tabular}{lcr}
\hline
$J/\psi$ decay   & $w_i$ (\%) 
                        & $ {\epsilon_i/\epsilon_{X}}-1$ (\%) 
                                        \\\hline
$e^+e^-$         &  5.9 & $+0.50\pm 0.43$ \\
$\mu^+\mu^-$     &  5.9 & $+0.92\pm 0.43$ \\
$\rho\pi$        &  2.1 & $+0.35\pm 0.65$ \\
$\gamma 3\pi^0$  &  1.1 & $+1.35\pm 0.62$ \\
$\omega\pi^0$    & 19.5 & $-0.15\pm 0.67$ \\
$2\pi^\pm3\pi^0$ &  6.4 & $+1.24\pm 0.86$ \\
$4\pi^\pm5\pi^0$ & 42.9 & $+0.10\pm 0.67$ \\
$6\pi^\pm2\pi^0$ & 14.4 & $-0.72\pm 0.80$ \\
$8\pi^\pm1\pi^0$ &  1.8 & $-0.67\pm 0.75$ \\
\hline
\end{tabular}
\end{center}
\end{table}

\begin{table}[thp]
\setlength{\tabcolsep}{0.4pc}
\catcode`?=\active \def?{\kern\digitwidth}
\caption{
Summary of ${\cal B}(J/\psi\to \ell^+\ell^-)$ results, showing 
numbers of $\psi(2S)\to\pi^+\pi^-\ell^+\ell^-$ decays, $N_{\diel}$ and $N_{\dimu}$;
efficiencies for observing those decays, $\epsilon_{\diel}$ and
$\epsilon_{\dimu}$; the corrected number
of such decays produced in the data sample; the number of
inclusive $\pi^+\pi^-J/\psi$ decays observed in the data sample
and extracted from the fit described in the text, $N_X$; the
corresponding efficiency, $\epsilon_X$, the corrected number of
inclusive $\pi^+\pi^-J/\psi$ decays produced in the data sample,
and relative statistical uncertainties on all quantities.
}
\label{tab:Sum}
\begin{center}
\begin{tabular}{lcc}
\hline
Quantity                      &  Value  & Rel. error (\%)\\ \hline
$N_{\diel}$                   &  14830  & 0.82 \\
$\epsilon_{\diel}$ (\%)       &  24.95  & 0.24 \\
$N_{\diel}/\epsilon_{\diel}$  &  59443  & 0.86\\
$N_{\dimu}$                   &  16697  & 0.77 \\
$\epsilon_{\dimu}$ (\%)       &  28.02  & 0.22\\
$N_{\dimu}/\epsilon_{\dimu}$  &  59588  & 0.81 \\
$N_X$                         & 395835  & 0.65 \\
$\epsilon_X$ (\%)             &  39.59  & 0.35 \\
$N_X/\epsilon_X$              &  999814 & 0.74\\
\hline
\end{tabular}
\end{center}
\end{table}

\end{document}